Reducing Latency and Noise in PPG-Based SpO$_2$ Measurements: A Kalman Filtering Approach Towards Acute Hypoxia Detection

Saud Lingawi, Garrett Frank, Benedictus H. Kartawidjaja, Mahsa Khalili, Brian Kwon, Calvin Kuo

**0. Abstract:** Photoplethysmography (PPG) is a common tool for monitoring cardiopulmonary health. Relying on absorption or reflectance of light by hemoglobin in the blood, the measured PPG waveform can be analyzed per heart beat using physiological assumptions to extract metrics ranging from heart rate to specific blood oxygenation (SpO$_2$). This has led to the widespread use of PPG for bedside clinical monitoring to wearable consumer health monitoring. However, PPG is notoriously noisy and the measured absorption or reflectance of light is sensitive to factors such as body movement and contact with the skin. To reduce the noise in the PPG-derived SpO$_2$, we developed combined traditional methods of estimating SpO$_2$ from the PPG waveform with a new method to extract changes in SpO$_2$ from the PPG waveform in a Kalman filter, and demonstrated its ability to better estimate SpO$_2$ in humans undergoing controlled hypoxia (down to 14% atmospheric oxygen). The Kalman filter reduced variability in SpO$_2$ to 4.30%SpO$_2$ compared to the beat-to-beat SpO$_2$ variability of 12.59%SpO$_2$. This mirrored current methods of window-averaging the beat-to-beat SpO$_2$, with a 30s window-average reducing SpO$_2$ variability to 4.73%. However, current window-average methods also introduce delays, with 10s and 30s window-averaging introducing delays of 5s and 14s respectively compared to the beat-to-beat SpO$_2$. The Kalman filter reduced this delay to within 3s of the beat-to-beat SpO$_2$, highlighting its ability to reduce noise while maintaining SpO$_2$ dynamics. This capability is particularly useful in reliably detecting clinically meaningful, but transient, hypoxic states, such as those observed during apnea.

# 1. Introduction

Photoplethysmography (PPG) sensors have long been used in clinical settings for the monitoring of vital signs such as blood oxygenation ($SpO_2$) and heart rate (HR) (1). These sensors optically measure changes in blood volume with the cardiac cycle through the absorption of light by hemoglobin in the blood. Operating non-invasively, they are a safe and robust measurement of such vital signs in acute emergency care (2). In more recent years, PPG sensors have become a standard in wearable devices for general physiological monitoring as well. With over 500 million wearables shipped annually, PPG-based devices, especially smartwatches, constitute a dominant share of consumer health wearables (3,4). In the commercial sector, these sensors are favored over other sensing modalities for several user-centred reasons (5). PPG-based wearable devices offer versatility and comfort, being adaptable to a range of different form factors that resemble commonly worn items (e.g., rings, watches) (6). Beyond that, they are also highly capable of performing continuous physiological monitoring without user input, making them suitable for accurate passive monitoring of several parameters such as HR, $SpO_2$, and respiratory rate (7).

In recent years, wearable device developers have started to shift from general wellbeing monitoring (e.g., step trackers, sleep quality monitoring) towards the detection of critical events. This is most notably seen in the development of fall detection algorithms using wearable movement sensors, and more recently atrial fibrillation detection algorithms using wearable single-lead electrocardiography (8–10). While wearable PPG sensors measure important physiological parameter to monitor for overall health and wellbeing, they can also provide an indicator of critical acute events. This is particularly true for $SpO_2$ in critical events where respiration is compromised and oxygenation may rapidly drop to physiologically dangerous levels. Obstructive sleep apnea and out-of-hospital cardiac arrest, respectively impacting roughly 1 billion and 4.5 million individuals annually, are prominent examples of acute deoxygenation as a result of cardiopulmonary dysfunction (11,12). In such events, the use of a non-invasive, rapid measurement of $SpO_2$ may aid in timely detection and intervention. Although commercial PPG-based devices are currently not validated for this purpose, PPG sensors are both suitable for such applications as well as adaptable to form factors that potential users would prefer (e.g., wristwatches and rings) (13,14).

However, PPGs are not without their flaws. Some commonly understood downfalls of the PPG are its sensitivity to noise and motion artifacts as well as differences in individual physiology (e.g., skin layer thickness and melanin) (15,16). To reduce the impact of signal noise on the accuracy of $SpO_2$ estimates, commercial developers do not estimate oxygenation at every heartbeat; instead, a window average over 10-30 seconds is taken (18–20). This introduces delays in estimating $SpO_2$, which may be detrimental to critical event detection. Delays up to 30 seconds in out-of-hospital cardiac arrest detection would reduce survivability (21), and most apnea events are shorter than 30 seconds (22), suggesting that a 10-30 second averaging window in $SpO_2$ may not respond quick enough or may even entirely miss a critical acute event (18).

Therefore, in order to rapidly detect acute changes in oxygenation, a methodology for $SpO_2$ estimation that reduces signal noise without window averaging is needed. As such, this paper focuses on the development of a Kalman filtering approach that reduces PPG signal noise while preserving physiological dynamics at the beat-level. Our approach relies on a novel mathematical model also based on the Beer-Lambert law that governs current SpO₂ estimation models, but developed with different assumptions to estimate changes in SpO₂ (dSpO₂). This model was tested in a benchtop hypoxia experiment, where participants were subjected to decreasing levels of oxygenation, to demonstrate both its ability to reduce beat-to-beat SpO₂ without introducing an estimation delay. This model will enable more robust and accurate PPG-based SpO₂ measurements to allow for the detection of acute critical conditions such as obstructive sleep apnea and out-of-hospital cardiac arrest.

## 2. Methods

*2.1 The Derivative Beer-Lambert Model*

PPG signals monitor the time-varying reflectance or absorption of the underlying tissue and vasculature at specific light wavelengths. While the PPG monitors a continuous signal, these are often evaluated per beat to extract SpO₂ measures. Mathematically, SpO₂ can be expressed as a ratio of the oxygenated hemoglobin concentration (HbO₂) to the total oxygenated and deoxygenated hemoglobin concentration in the blood (HbO₂ + Hb):

$$SpO_2 = [\mathbf{HbO_2}] / ([\mathbf{HbO_2}] + [\mathbf{Hb}])$$

Calculating SpO₂ using PPG sensors relies on mathematical manipulations of the Beer-Lambert law, which relates the absorption of light (A) by a medium to: the concentration ([c]) of the absorbing species; its intrinsic absorptivity (ε, empirically determined); and the light path length (L):

$$A = \varepsilon L [c]$$

Such that:

$$A_{HbO2} = \epsilon_{HbO2} L [\mathbf{HbO_2}] \; and \; A_{Hb} = \epsilon_{Hb} L [\mathbf{Hb}]$$

However, a critical consideration when using absorption as a method of quantifying concentrations is that the emitted light is not only absorbed by the species of interest in the capillary network, or:

$$A_{PPG} = A_{hemoglobin} + A_{tissue}$$

$$A_{Hemoglobin} = \epsilon_{HbO2} L [\mathbf{HbO_2}] + \epsilon_{Hb} L [\mathbf{Hb}]$$

Therefore, isolating the contribution of hemoglobin to the absorption of PPG light requires the incorporation of two wavelengths of light with different absorption characteristics. Typically, red (~660 nm) and infrared (IR, ~940 nm) wavelengths are used such that:

$$A_{IR} = \varepsilon_{HbO2}^{IR} L [\mathbf{HbO_2}] + \varepsilon_{Hb}^{IR} L [\mathbf{Hb}] + A_{tissue}^{IR}$$

$$A_{Red} = \varepsilon_{HbO2}^{red} L [\mathbf{HbO_2}] + \varepsilon_{Hb}^{red} L [\mathbf{Hb}] + A_{tissue}^{red}$$

The current method for interpreting PPG data to extract SpO2 relies on several assumptions: the light path (L) changes through the heart beat with the expansion and contraction of blood vessels; the concentrations of $[Hb]$ and $[HbO_2]$ remain constant through the heart beat; and the absorption from other tissue ($A_{tissue}^{red}$ and $A_{tissue}^{IR}$) remains constant through the heart beat. Thus, calculating the relative absorption of red and IR light allows for the derivation of a unitless ratio of ratios, $R$, that relies on both the pulsatile absorption (AC) and constant absorption (DC).

$$R = \frac{(AC/DC)^{Red}}{(AC/DC)^{IR}}$$

After which SpO2 can be calculated as a measure of $R$ and the intrinsic absorptivity coefficients ($\varepsilon$) of Hb and HbO2 in red and IR light:

$$SpO_2 = \left( \frac{-\varepsilon_{Hb}^{red} + R * \varepsilon_{Hb}^{IR}}{-R * (\varepsilon_{HbO2}^{IR} - \varepsilon_{Hb}^{IR}) + (\varepsilon_{Hb}^{red} - \varepsilon_{HbO2}^{red})} \right) * 100$$

While the ratio of ratios method is the standard for estimating SpO2 from the PPG waveform, the reliance on these key assumptions makes it susceptible to noise across beats. To develop our model, we relied on a different set of assumptions. First, we assume the concentrations of $[HbO_2]$ and $[Hb]$ do not remain constant, but change in an equal and opposite manner (i.e., the total concentration of hemoglobin remains constant) between beats. This is particularly true in clinical applications such as acute hypoxia, desaturation, and pulselessness. We also consider that, at the same point in across heart beats, that the path length L is assumed not to change. This allows for the changes in light absorption to be expressed as changes in concentration, rather than changes in path length. Finally, we assume that absorption from the tissue is correlated with the hemoglobin concentration. Together, this allows us to express absorption as:

$$\Delta[HbO_2] = -\Delta[Hb] \gg \Delta L \approx 0, \text{ and } \Delta A_{tissue} \approx \beta \Delta[HbO_2]$$

$$\Delta A_{IR} = \varepsilon_{HbO2}^{IR} L \Delta[HbO_2] - \varepsilon_{Hb}^{IR} L \Delta[HbO_2] + \beta^{IR} \Delta[HbO_2]$$

$$\Delta A_{Red} = \varepsilon_{HbO2}^{red} L \Delta[HbO_2] - \varepsilon_{Hb}^{red} L \Delta[HbO_2] + \beta^{red} \Delta[HbO_2]$$

Simplifying these assumptions leads to a more direct relationship between these changes:

$$\Delta A_{IR} = \Delta[HbO_2](\varepsilon_{HbO2}^{IR} L - \varepsilon_{Hb}^{IR} L + \beta^{IR}) = \gamma^{IR} \Delta[HbO_2]$$

$$\Delta A_{red} = \Delta[HbO_2](\varepsilon_{HbO2}^{red} L - \varepsilon_{Hb}^{IR} L + \beta^{red}) = \gamma^{red} \Delta[HbO_2]$$

$$\Delta Diff: \Delta A_{IR} - \Delta A_{red} = (\gamma^{IR} - \gamma^{red}) \Delta[HbO_2]$$

This the allows us to express changes in SpO2 as correlated with the difference in change in absorptions:

$$\Delta SpO_2 = \Delta[HbO_2] / ([HbO_2] + [Hb])$$

$$\therefore \Delta SpO_2 \propto \Delta Diff \text{ or } \Delta SpO_2 \approx \alpha \Delta Diff$$

Where $\alpha$ is a calibrated constant relating $\Delta SpO_2$ to $\Delta Diff$, the difference between derivative IR and red PPG light absorption ($\Delta A_{IR} - \Delta A_{red}$). This modified derivative model then theoretically allows for the direct estimation of $\Delta SpO_2$, which is particularly relevant for clinical events such as acute hypoxia, desaturation, and pulselessness to measure sudden changes in oxygenation.

*2.2 The Low-Latency Beat-to-Beat SpO₂ Multivariate Kalman filter*

Thus, we now have two different formulations of the PPG signal to obtain an SpO₂ signal that is often noisy, as well as $\Delta SpO_2$ which can be integrated but will likely be susceptible to integration drift. Utilizing both pieces of information is a classic application for Kalman filters, mirroring work in vehicle dead reckoning and IMU orientation. Fundamentally, the purpose of a Kalman filter (*Table 1*) is to estimate the true value of a *state* ($\vec{x}_k$) from sensor *measurements* ($\vec{z}_k$) at a particular point in time $k$. For our Kalman filter, $k$ refers to a beat identified in the PPG signal. The sensor measurements are related to the state through an *observation model* (described by the time-invariant observation matrix $H$) as the state evolves over time through a *dynamic system* (described by the time-invariant state transition matrix $F$). The key to the Kalman filter is that both the state and measurements are assumed to have error, described by covariance matrices $P$ and $R$ for the state and measurement respectively. We also further introduce process noise $Q$ on the state as it evolves through time. These covariances are used to form a minimum variance *a posteriori* estimate of the state ($x_{k|k}$) and its covariance ($P_{k|k}$) through a Kalman gain ($K$). This Kalman gain ($K$) weights contributions from an *a priori* state estimate ($x_{k|k-1}$) and covariance ($P_{k|k-1}$) from the state dynamics (prediction step) and the current sensor measurements ($z_k$) and its covariance ($R$) (update step).

*Table 1. Summary of the Kalman Filter Equations and Matrices*

| Stage | Equations | Matrix | Name | Value |
|---|---|---|---|---|
| Predict | $x_{k\|k-1} = Fx_{k-1\|k-1}$<br>$P_{k\|k-1} = FP_{k-1\|k-1}F^T + Q$ | $F$ | State transition | $\begin{bmatrix} 1 & dt \\ 0 & 1 \end{bmatrix}$ |
|  |  | $Q$ | Process noise | $\begin{bmatrix} 0.01 & 0 \\ 0 & 0.01 \end{bmatrix}$ |
| Update | $K_k = P_{k\|k-1}H^T(HP_{k\|k-1}H^T + R)^{-1}$<br>$x_{k\|k} = x_{k\|k-1} + K_k(z_k - Hx_{k\|k-1})$<br>$P_{k\|k} = (I - K_kH)P_{k\|k-1}$ | $H$ | Observation | $\begin{bmatrix} 1 & 0 \\ 0 & 1/\alpha \end{bmatrix}$ |
|  |  | $R$ | Measurement noise | $\begin{bmatrix} 4 & 0 \\ 0 & 25 \end{bmatrix}$ |

The state transition matrix ($F$) describes a simple, first order system integrating the *a posteriori* $\Delta SpO_2$ to get the *a priori* estimate of $SpO_2$. In the observation matrix ($H$), we directly measure $SpO_2$ in the state using the "ratio of ratios" algorithm in the PPG signal. For the state $\Delta SpO_2$, this is related to the measurement of $dDiff$ through the scaling factor $\alpha$ derived previously in the PPG signal. In our formulation, we assume the process noise ($Q$) and measurement noise ($R$) are time-invariant (e.g. these systems always have a baseline level of noise present). Because the dynamic system is a simple first order integration, we assumed the process noise to be small compared to other sources.

*2.3 Benchtop Hypoxia Experimental Design*

To validate this Kalman filter approach, we conducted a controlled benchtop hypoxia study using healthy adult participants under varying inspired oxygen conditions. This study was approved by the University of British Columbia Research Ethics Board (H24-00763). The inclusion criteria were as follows: 1) participant must be 19-40 years old, and 2) participant must be able to provide consent in English. Participants were recruited on a semi-convenience basis, enrolling any participant willing to partake in the experiment while ensuring a representative distribution of gender and skin type. The Fitzpatrick skin type scale was used to ensure a broad representation of different skin tones, particularly important when evaluating light absorption-based technologies wherein light absorption has been suggested to operate differently when interacting with varying levels of melanin. Each experiment comprised of three sessions, spaced out a minimum of 24 hours apart. We conducted three sessions per participant to evaluate the intra-participant agreement in $SpO_2$ estimations and $\alpha$ approximations.

Participants were instrumented with a custom PPG sensor (MAXREFDES117, Maxim Integrated, USA) on the left middle finger, connected to an Adalogger M0 microcontroller (Adafruit Industries, USA), and a MightySat pulse oximeter (Masimo, USA) on the left index finger. Each 25-minute session included five 5-minute blocks: baseline (21% $FiO_2$), hypoxia induction (16%, 14%, and 12% $FiO_2$ via airbag), and recovery (21% $FiO_2$). Participants breathed through a sealed mouthpiece with nose clamping to ensure precise $FiO_2$ control.

*2.4 Data Analysis*

All data analysis was performed using Python 3.12 and MATLAB R2022a.

*2.4.1 PPG and Masimo Time Synchronization*

The length of each session was marked by two conditions: 1) real clock time synchronized with the Masimo system, and 2) induced large motion artifacts in the raw data from the custom PPG corresponding to the end of the session. Using MATLAB, custom PPG-derived signals (sampled at 50 Hz) were temporarily downsampled and interpolated to align with lower-rate Masimo $SpO_2$ timestamps (sampled at 1 Hz). This allowed for alignment and clipping based on both the Masimo real clock time and custom PPG motion artifacts, after which custom PPG-derived signals were resampled to 50 Hz.

*2.4.2 PPG Beat Detection*

Using MATLAB, raw, unfiltered, time-synchronized PPG-derived red and IR signals were first bandpass filtered (4$^{th}$-order Butterworth filter, frequency band: 0.5-3.5 Hz) to isolate the fundamental frequency of the heart and its harmonics. Both red and IR bandpass filtered signals were run through a peak detection function with a minimum peak distance of 0.5s. For PPG trough detection, the signal was simply inverted and run through an identical analysis, after which peaks and troughs were combined and mapped to the lowpass filtered signals.

All subsequent analyses were done in Python. A custom function was written to ensure that detected beats represent true peak-trough pairs by enforcing a maximum time difference of 0.7s between each peak and its trough. Another custom function was then written to ensure that both red and IR signals were aligned in their detected beats by enforcing a maximum time difference of 0.05s between beats in the red and IR signals (selected as a strict constraint for a semi-lagless multi-channel PPG device per manufacturer specifications). Prior to further analysis, beat alignment was assessed by comparing IR PPG-derived heart rate (HR) against the reference Masimo HR. IR PPG-derived HR was computed from the inverse of RR intervals between beat centers, and compared against interpolated Masimo HR values at the same timestamps. Agreement was assessed using root mean square error (RMSE) as well as a Bland-Altman analysis to assess systematic bias and limits of agreement.

### 2.4.3 SpO$_2$ Calculation

Going back to the raw, unfiltered, and time-sycnrhonized PPG signals, we this time lowpass filtered (4$^{th}$-order Butterworth filter, cutoff frequency = 3.5 Hz) the red and IR signals to remove any high frequency noise or artifact. We did not perform the bandpass as the DC component of the red and IR signals are required for estimating SpO$_2$ through the ratio-of-ratios method. SpO$_2$ was calculated using the standard ratio-of-ratios method for three different conditions: 1) beat-to-beat SpO$_2$, 2) 10-second moving window-averaged SpO$_2$, and 3) 30-second moving window-averaged SpO$_2$. The time-averaged SpO$_2$ conditions were computed every 1s, mirroring common techniques for commercial SpO$_2$ analysis. Pearson correlation coefficients were calculated for each SpO$_2$ condition compared to the reference Masimo to assess SpO$_2$ agreement. To assess the lag induced by the window-averaging, a time-lagged cross-correlation was performed on both the 10-second and 30-second window-averaged SpO$_2$ compared to the beat-to-beat SpO$_2$, taking the lag at the maximal Pearson correlation coefficient.

### 2.4.4 $\Delta Diff$ Calculation, $\alpha$ Optimization, and Kalman filter comparison

To incorporate our new model, we first had to calibrate the parameter $\alpha$. Using the NuMPY library, the numerical derivative of the following signals was calculated via finite differences: 1) red PPG peaks ($\Delta A_{red}$), 2) IR PPG peaks ($\Delta A_{IR}$), and 3) beat-to-beat SpO$_2$ ($\Delta SpO_2$). $\Delta Diff$ was computed as the difference between peak $\Delta A_{IR}$ and $\Delta A_{red}$ signals and compared to $\Delta SpO_2$. Prior to computing each session's $\alpha$ value, the $\Delta Diff$ and $\Delta SpO_2$ signals were lowpass filtered (4$^{th}$-order Butterworth filter, cutoff frequency = 0.1 Hz) in order to isolate SpO$_2$ dynamics from noisy signals. Then, each session's $\alpha$ value was computed through least squares error minimization that assumes a linear relationship between $\Delta SpO_2$ and $\Delta Diff$. These $\alpha$ were considered each session's *local* $\alpha$ value, as in the value derived from the session itself. For *n* sessions, *n local* $\alpha$ values were computed.

Two other $\alpha$ values were computed for further analysis and optimization: 1) the *local median* $\alpha$ defined as the median of the *local* $\alpha$ values across all sessions of a participant (i.e., *m local median* $\alpha$ values computed for *m* participants), and 2) the *global* $\alpha$ defined as the median of the *local* $\alpha$ values (i.e., 1 *global* $\alpha$ for all *n local* $\alpha$ values).

Finally, the Kalman filter was constructed according to the equations and matrices summarized in *Table 1*. The filter was initialized with an estimate of 95% and 0 for SpO$_2$ and dDiff, respectively, and a 2x2 identity matrix for the error covariance matrix ($P_0$). For each session, the Kalman filter was constructed and applied three times in parallel, once for each $\alpha$ value.

To compare the performance of the different $\alpha$ values in tracking $SpO_2$ dynamics, the Pearson correlation coefficient (r) between Kalman filter $SpO_2$ estimates and the reference Masimo $SpO_2$ measurements were calculated. To evaluate statistical significance, the Friedman non-parametric statistical test was run on all possible $\alpha$ groupings. If a significant difference was found, a subsequent Wilcoxon signed-rank statistical test with a Holm-Bonferroni correction was run to identify the pairwise statistical significance between $\alpha$ values. This was to assess the sensitivity of $\alpha$ to each individual and assess whether we could use a global $\alpha$. The KF with the highest correlation to Masimo $SpO_2$ was then compared to the beat-to-beat, 10s window-averaged, and 30s window-averaged $SpO_2$ and a similar time-lagged cross-correlation was conducted to assess its responsiveness in comparison to the beat-to-beat $SpO_2$.

To further evaluate the performance of the Kalman filter in reducing signal variance and producing accurate $SpO_2$ estimates, root-mean-squared errors (RMSE) were calculated between beat-to-beat PPG $SpO_2$, KF $SpO_2$, and Masimo $SpO_2$. Additionally, standard deviations (SD) were calculated and compared for each of the 5 breathing conditions (baseline 21% $FiO_2$, 16% $FiO_2$, 14% $FiO_2$, 12% $FiO_2$, and recovery 21% $FiO_2$).

## 3. Results

### 3.1 Participant Summary

The benchtop hypoxia experiment was completed by 13 participants, with a total of 38 sessions. All but two completed three sessions each, with one completing four sessions and one withdrawing from the experiment after one session. Of the 38 sessions, 3 sessions were left out of the analysis due to incomplete data or sensor errors from the custom PPG (n=35). One additional session was missing the reference Masimo data, for which analyses were limited to comparisons between the Kalman filter and custom PPG-derived $SpO_2$.

*Table 2. Summary of the Benchtop Hypoxia Experiment*

| Age | $29.08 \pm 4.97$ | |
|---|---|---|
| **Gender** | Women | 7 |
| | Men | 6 |
| **Fitzpatrick Skin Type** | I | 0 |
| | II | 6 |
| | III | 4 |
| | IV | 0 |
| | V | 2 |
| | VI | 1 |
| **Total Sessions** | 38 | |
| **Sessions Analyzed** | 35 | |

## 3.2 PPG Beat Alignment & SpO$_2$ Calculation

After performing beat alignment and comparing IR PPG-derived HR to the interpolated Masimo HR, we found a global RMSE of 15.95 ± 12.93 BPM. Overall, IR PPG-derived HR showed good agreement with interpolated Masimo HR, with a small bias of -5.20 bpm centred around 0 and limits of agreement ranging from -34.75 to 24.35 bpm. Notably, the Bland-Altman analysis revealed that the majority of HR comparisons lie within the limits of agreement, with comparisons outside these limits likely being attributed to sensitivity of beat-to-beat HR calculations in the custom PPG to noise.

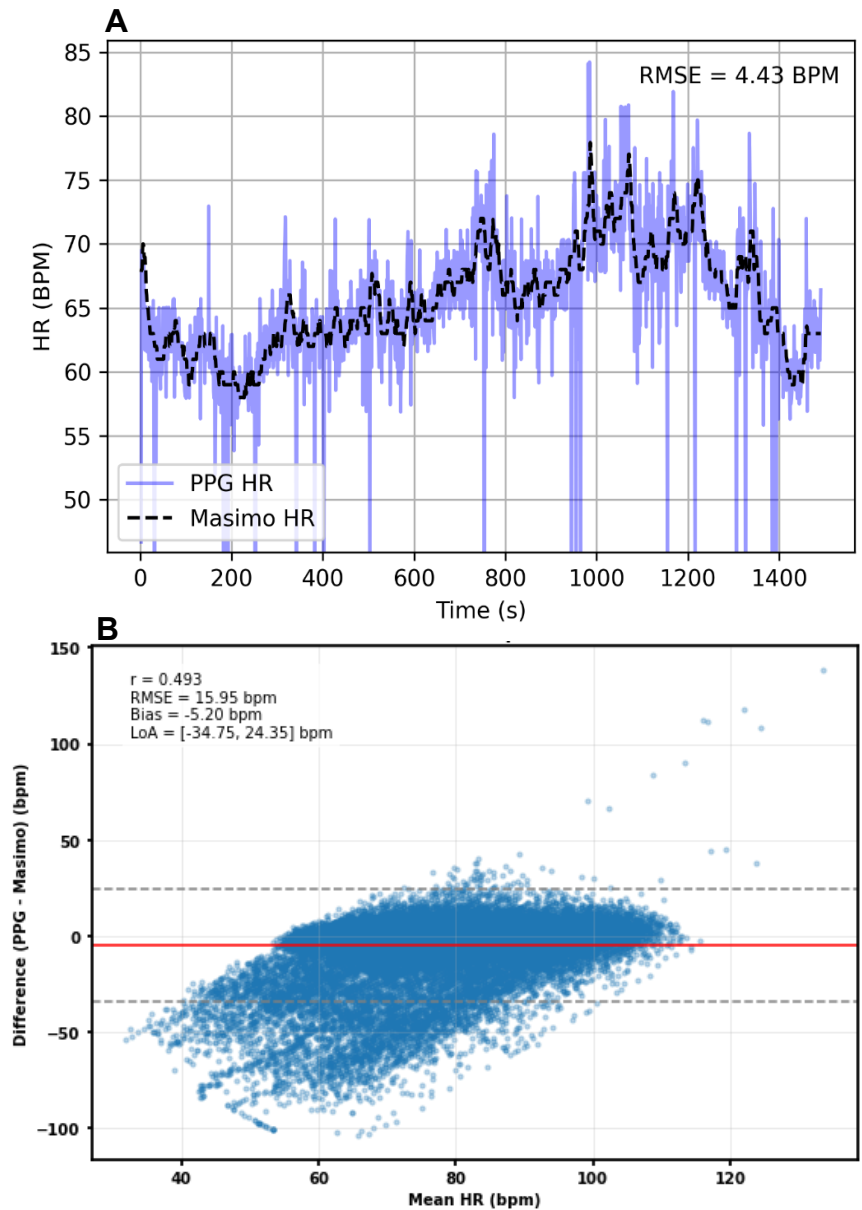

*Figure 1. (**A**) Representative Comparison of Custom PPG-Derived HR and Interpolated Masimo HR; (**B**) Bland-Altman Analysis of Custom PPG-Derived HR and Interpolated Masimo HR*

Comparing custom PPG-derived SpO$_2$ to the Masimo SpO$_2$ reveals a low median correlation coefficient of 0.258 between beat-to-beat SpO$_2$ and Masimo SpO$_2$. Median correlation increases with increasing window averages, with the 10s window-average and 30s window-average exhibiting median correlations of 0.502 and 0.738, respectively. Despite the low correlation between beat-to-beat SpO$_2$ and Masimo SpO$_2$, beat-to-beat SpO$_2$ still successfully tracks the oxygenation dynamics exhibited by Masimo SpO$_2$ across breathing conditions. This suggests that this low correlation is primarily due to noise at the beat resolution. Averaging SpO$_2$ over larger windows (i.e., 10-30s) reduces the RMSE between the associated PPG-derived SpO$_2$ and Masimo SpO$_2$, with mean RMSE values being highest for the beat-to-beat SpO$_2$ (11.94 ± 7.87) and lowest for the 30s averaged SpO$_2$ (6.51 ± 7.64). The time-lagged cross-correlation analysis revealed that, compared to the beat-to-beat SpO$_2$, the 10s and 30s averaged SpO$_2$ exhibit median time lags of 5 and 14s, respectively.

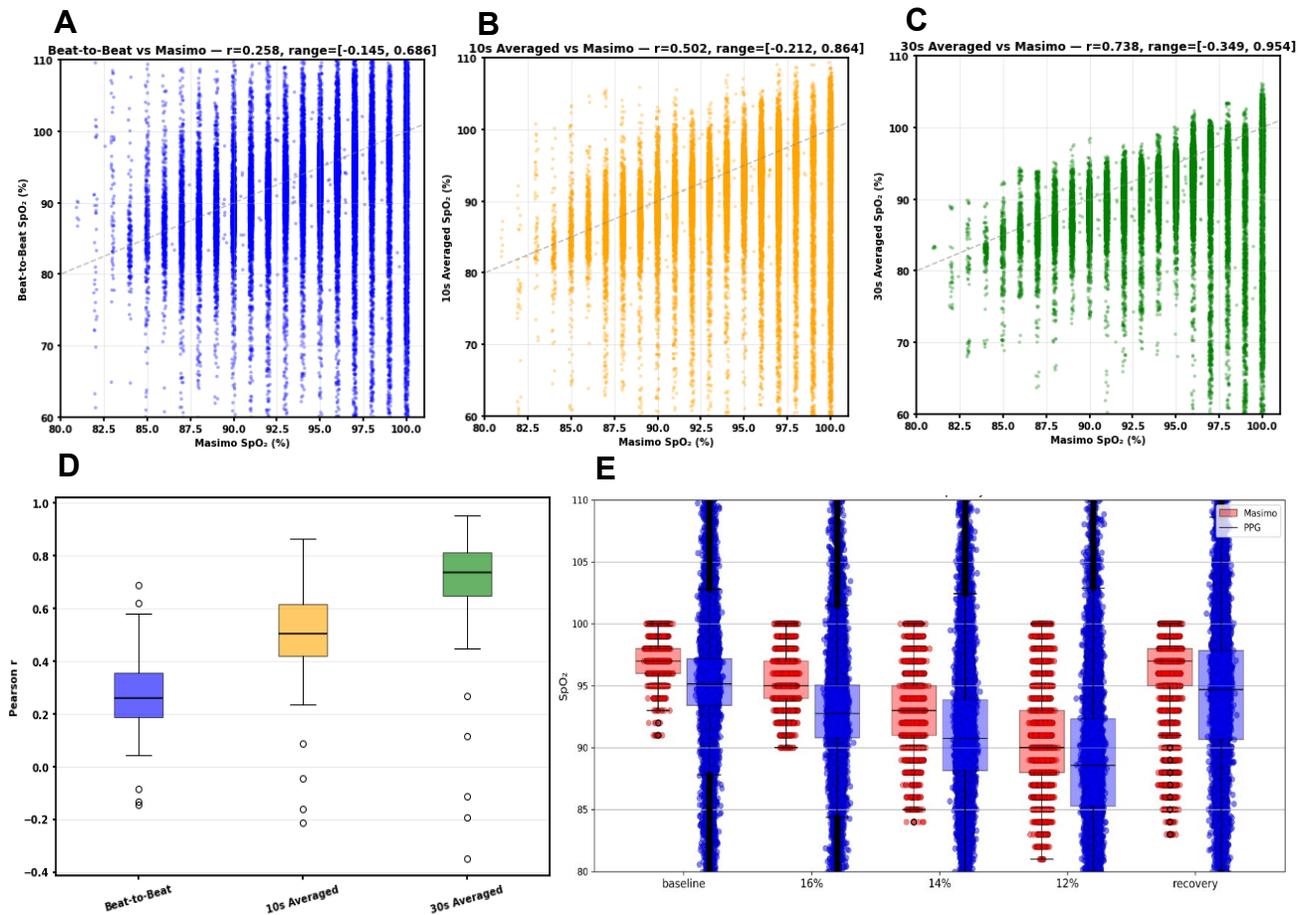

*Figure 2. (**A-C**) Correlation of Beat-to-Beat, 10s Averaged, and 30s Averaged SpO$_2$ to Masimo SpO$_2$; (**D**) Boxplot Summary Illustrating the Range and Median Correlation Between SpO$_2$ Conditions and Masimo SpO$_2$; (**E**) Summary of Overall Spread of Beat-to-Beat and Masimo SpO$_2$ Values Across Breathing Conditions*

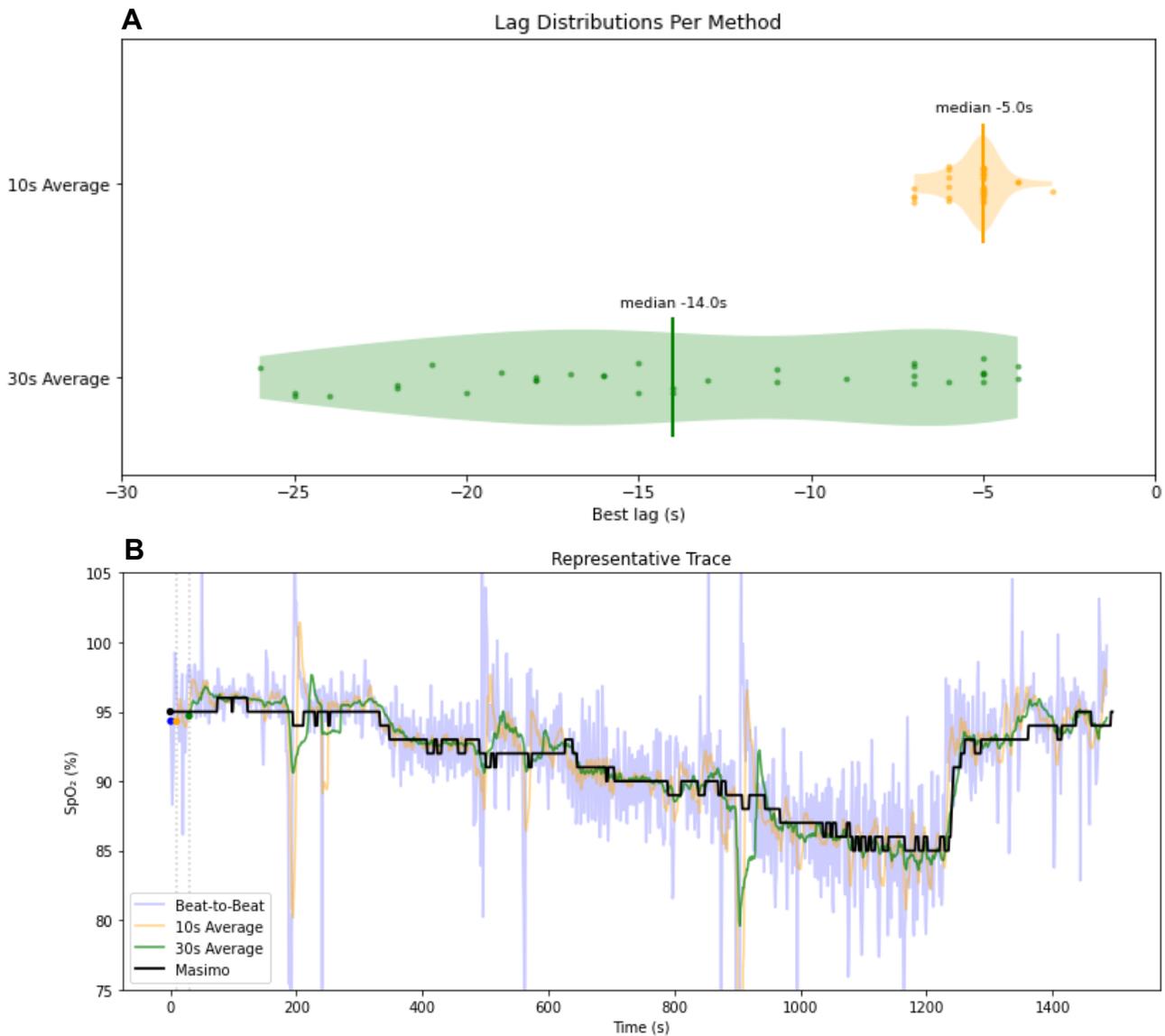

*Figure 3. **(A)** Time-Lagged Cross-Correlation of the 10s and 30s Averaged SpO2, Compared to Beat-to-Beat SpO$_2$; **(B)** Representative Trace Illustrating SpO$_2$ Tracking of the Beat-to-Beat, 10s Averaged, and 30s Averaged Methods Compared to Masimo SpO$_2$*

*3.3 Alpha Optimization & Kalman Filter Performance*

The *global* $\alpha$ resulted in the highest correlation to Masimo SpO$_2$ in 51.43% of sessions. However, performing the Friedman test to compare $\alpha$ types did not reveal a significant difference in their correlations to the Masimo SpO$_2$. The absence of statistical significance among $\alpha$ types paired with the *global* $\alpha$ having the highest correlation to Masimo SpO$_2$ suggests that the *global* $\alpha$ is the most suitable $\alpha$ value to use across sessions.

This *global* $\alpha$ produced a median correlation of 0.687, outperforming both the beat-to-beat and 10s averaged SpO$_2$, as well as more closely resembling the 30s averaged SpO$_2$. The time-lagged cross correlation against the beat-to-beat SpO₂ also revealed a median lag of 3s, an improvement from both the 10s and 30s averaged SpO₂. This supports that the KF successfully reduces beat-level noise while still preserving beat-level dynamic tracking.

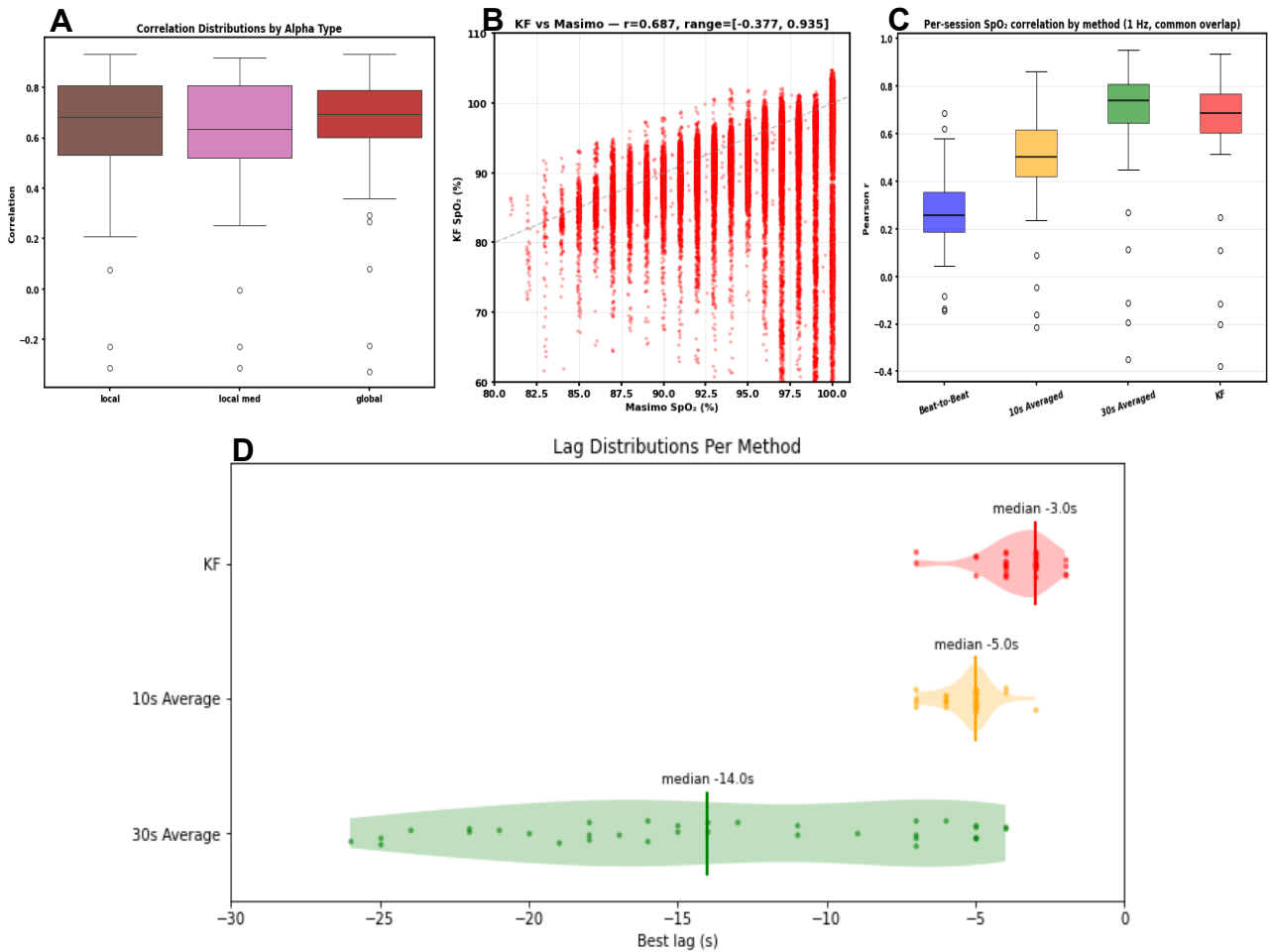

Figure 4. *(A) Boxplot Summary of Overall Spread of Correlation Between Kalman Filter outputs and Masimo SpO₂ Across Different α Values; (B) Correlation of the Global α KF SpO₂ to Masimo SpO₂; (C) Boxplot Summary Comparing the Range and Median Correlation Between SpO₂ Conditions and Masimo SpO₂; (D) Time-Lagged Cross-Correlation of the Global α KF SpO₂ Compared to the 10s and 30s Averaged SpO₂*

As shown in the representative trace below, the KF output (red) closely tracks oxygenation changes in response to each inspired FiO₂ transition (vertical lines), including declines to 16%, 14%, and 12% oxygen and subsequent reoxygenation. Notably, while the KF reduces noise in beat-to-beat measurements, it also still preserves beat-level responsiveness that is lost window-averaged estimates. When comparing the performance of the KF in tracking the dynamics exhibited by the Masimo SpO₂, we found that the KF reduces the RMSE from 11.94 ± 7.87 %SpO₂ in the beat-to-beat PPG SpO₂ to 5.55 ± 5.13 %SpO₂. This was a greater reduction in RMSE compared to the 10s window-average (7.75 ± 6.32 %SpO₂) and similar to the RMSE for the 30s window-average (5.82 ± 5.87 %SpO₂). Additionally, while the variability range of beat-to-beat SpO₂ was ±9.59% to ±12.59% across breathing conditions, the variability range in KF SpO₂ dropped to ±2.75% to ±4.30%. Again, this was a greater reduction in variability compared to the 10s window-average (±5.09% to ±7.01%) and similar to the 30s window-average (±2.80% to ±4.73%). For comparison, the Massimo system had a ±0.55% to ±2.91% variability range. This illustrates a substantial 68.9-75.3% reduction in variability in the KF output compared to the beat-to-beat PPG SpO₂.

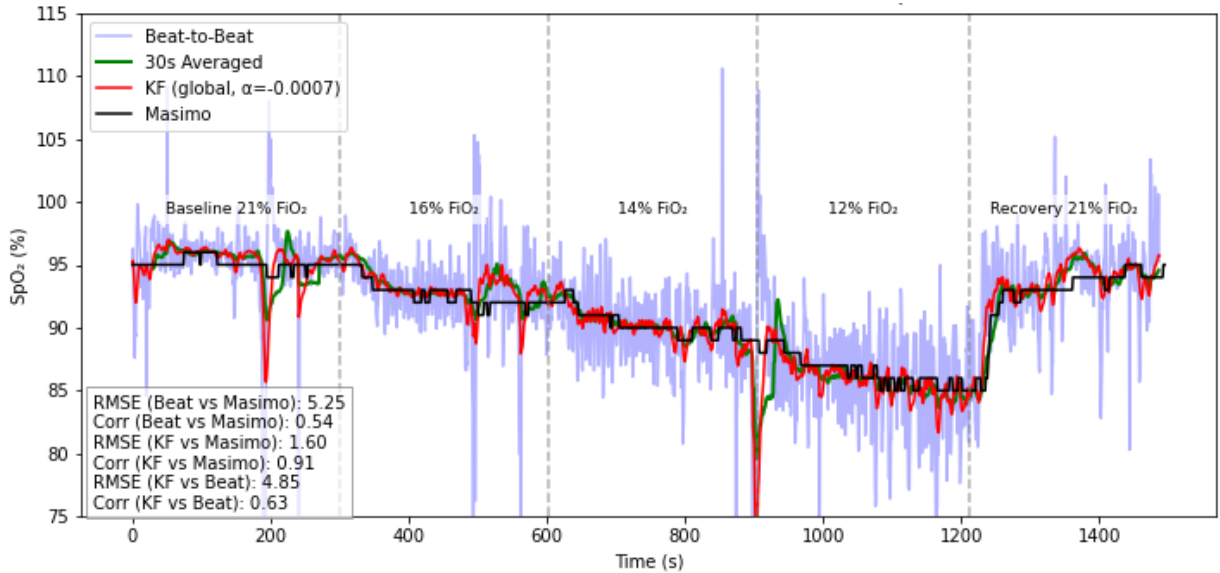

*Figure 5. Sample KF Implementation Using the Global α*

**Discussion:**

Remote detection of acute critical illness events in real-time requires a lag-less sensor with robust measurements that are capable of performing state estimations under pathological conditions (e.g., acute loss of pulse). In this paper, we detailed a Kalman filtering approach that reduces the variability of beat-to-beat SpO₂ measurements, eliminating lag introduced by window-averaging over longer periods of time to obtain a stable measurement. This is achieved through a novel parameter in SpO₂ estimation, $\Delta Diff$, that allows for a $\Delta SpO_2$ measurement through the tunable parameter $\alpha$.

A key feature of this KF approach is the preservation of responsiveness to acute changes. As can be seen in the representative trace in *Figure 4*, while the KF produces a signal of lower variance compared to the beat-to-beat SpO₂ signal, spikes in SpO₂ are still captured. In the case of noise-induced spikes, such as the sharp spikes observed during baseline and the transition from 14 to 12% FiO₂, the KF quickly returns to a low-variance signal while reducing the magnitude of the faulty peak. Such behavior confirms that the dynamic model built into the KF is strong enough to quickly readjust KF predictions and not get derailed by a noisy signal. However, in the case of true rapid spikes in SpO₂, as seen during the transition from 12% FiO₂ to recovery, the KF successfully rapidly adjusts to and stabilizes at the higher SpO₂ values without over- or under-estimating the true state.

One difficulty of using the new mathematical derivation for $\Delta SpO_2$ is in the choice of a calibration parameter $\alpha$. Given the mathematical origination of $\alpha$ capturing elements of tissue absorption, this parameter should theoretically require re-calibration for each individual and potentially each session, making the model impractical. However, our comparison of *local*, *local median*, and *global* $\alpha$ values revealed the Kalman Filter was not sensitivity to the calibration parameter $\alpha$. Interestingly, although the performances of the different $\alpha$ values were not statistically different, the *global* $\alpha$ did produce the highest correlation to Masimo $SpO_2$ compared to the other two $\alpha$ values. This suggests that, while each session resulted in a different $\alpha$ value, the *global* $\alpha$ could still be used across all sessions and would result in a KF implementation that successfully reduces beat-to-beat variability while preserving beat-to-beat dynamics. Moreover and more importantly, this suggests that *global* $\alpha$ is generalizable to different users in different use cases, eliminating the need for person-specific calibration and further strengthening the empirical relationship between $\Delta Diff$ and $\Delta SpO_2$.

There are a handful of limitations to the work detailed in this paper. Firstly, the age range of the inclusion criteria limited participants to those who are less likely to experience acute desaturations or pulselessness in a real-world setting. This age range was selected in accordance with the ethics committee to avoid potential physiological complications associated with exposing individuals to lower amounts of Oxygen. Secondly, our current implementation of $\Delta Diff$ was computed between beats in order to match the resolution of beat-to-beat $SpO_2$ and simplify the KF implementation. While indeed simplifying the KF implementation, this approach introduces more complex pre-processing and still relies on the presence of a pulse. Further work will implement the integration of a continuous $\Delta Diff$ signal, potentially allowing for estimation of $SpO_2$ in pulseless conditions. Finally, these experiments were conducted on motionless participants in highly controlled settings, therefore the current KF performance does not account for sensor noise introduced by motion and/or changing environmental conditions.

Conventional commercial wearable devices rely on averaging physiological parameters over 10-20s windows to produce accurate estimates, which reduces high frequency noise but introduces delays that can result in missing acute critical health events. The Kalman filtering approach in this paper relying on a modified Beer-Lambert model, provides an alternative to window-averaging that can produce accurate state estimates and preserve acute changes in physiological dynamics. Such a model has potential implications for the timely detection of critical illness events such as acute hypoxia, desaturation, or pulselessness.